\documentclass{slides}

\usepackage{epsf}

\begin{document}

\begin{slide}       
\vspace*{-1in}
\begin{center}
{\bf  AN EFFECTIVE LAGRANGIAN APPROACH TO NUCLEI}
\end{center}

\medskip
\begin{flushright}
   \small\it
   \makebox[4.0in][l]{\null R. J. Furnstahl and Hua-Bin Tang}
   \makebox[4.0in][l]{\null The Ohio State University}
   \makebox[4.0in][l]{\null {\rm (}in collaboration
                                   with B. D. Serot{\rm )}}
\end{flushright}

\vspace*{.5in}
\raggedright
\normalsize
{\bf Abstract:}
\small
\begin{quote}
The construction of a general effective lagrangian
consistent with the symmetries of QCD
and intended for applications to finite-density systems is
discussed. The low-energy structure of the composite nucleon is
described within the theory from vector-meson dominance.
Results are given for finite nuclei
and nuclear matter
at one-loop order. All of the coupling constants, appropriately defined
according to naive dimensonal analysis, are found  to be
natural.
\end{quote}

\normalsize
{\bf Outline:}
\begin{itemize}
   \vspace*{-.3in}
 \item Introduction
   \vspace*{-.5in}
 \item Physical Ingredients
   \vspace*{-.5in}
 \item The Lagrangian
   \vspace*{-.5in}
 \item EM Structure of the Nucleon
   \vspace*{-.5in}
 \item
      One-Loop Results
  \vspace*{-.5in}
 \item
      Summary and Future Work
\end{itemize}

\end{slide}

\begin{slide}       
\vspace*{-1in}
\centerline{{\bf INTRODUCTION}}
\raggedright
\begin{itemize}
\normalsize
 \item
\vspace*{-.1in}
   Why adopt the approach of  an effective hadronic lagrangian?
   \small
   \begin{itemize}
      \item Modern renormalization theory makes sense of
            nonrenormalizable effective lagrangians.
\vspace*{-.2in}
      \item QCD at low energies
      is equivalent to some effective theories with
        low-energy degrees of freedom.
\vspace*{-.2in}
      \item QHD mean-field theory is successful for
            nuclei; it is thus interesting to study nuclei from
            the modern viewpoint of effective theories.
\vspace*{-.2in}
      \item The composite structure of the nucleon can be described
            in increasing detail by adding additional
            nonrenormalizable interactions.
   \end{itemize}
\normalsize
\vspace*{-.2in}
 \item
   Why include non-Goldstone bosons?
   \small
   \begin{itemize}
      \item The $E^4$ low-energy constants of ChPT are saturated by
            vector-meson contributions.
\vspace*{-.2in}
      \item The important intermediate-range
            NN interactions are efficiently described by their
            exchange ($\sigma$, $\omega$, $\cdots$).
\vspace*{-.2in}
      \item Large $N_c$ QCD reduces to a nonlinear effective theory
            of mesons with the nucleon emerging as a soliton.
\vspace*{-.2in}
      \item At finite density, nucleon bilinears such as
            $\overline{N}N$ and $N^\dagger N$ develop expectation
            values that are conveniently and efficiently described by
            the appropriate meson mean fields.
   \end{itemize}
\end{itemize}
\end{slide}

\begin{slide}       
\vspace*{-1in}
\centerline{{\bf PHYSICAL INGREDIENTS}}
\raggedright
\begin{itemize}
\normalsize
 \item
\vspace*{-.2in}
  Chiral Symmetry:\ \ \
\small
     Use CCWZ's nonlinear realization of $SU(2)_{\rm L} \otimes
SU(2)_{\rm R}$ symmetry.
\normalsize
   \vspace*{-.5in}
 \item Gauge Invariance:\ \ \
\small
   Allow for general gauge invariant terms.
\normalsize
   \vspace*{-.5in}
  \item
      Broken Scale Invariance:\ \ \
\small
     A light scalar {\it might\/}
     be associated with the broken scale invariance
    of QCD.
\normalsize
   \vspace*{-.5in}
   \item
      NDA and Naturalness:\ \ \
\small
  Naive dimensional analysis (NDA) and a ``naturalness'' assumption
  justify a truncation of the lagrangian. They are useful as
  long as the density is not too high.

\normalsize
  \vspace*{-.5in}
   \item
      Field Redefinitions:\ \ \
\small
     Use to simplify the interaction terms
 between the nucleons and the non-Goldstone bosons.

\normalsize
  \vspace*{-.5in}
   \item
     Vacuum Effects:\ \ \
\small
           Vacuum baryon and non-Goldstone boson loops
are ``integrated out,'' with their effects buried in the
remaining coefficients.
This disentangles vacuum dynamics from valence nucleon
dynamics.

 \end{itemize}

\end{slide}

\begin{slide}       
\vspace*{-1in}
\normalsize
\centerline{$\bullet$ {\bf Chiral Symmetry}}
\raggedright

\small
\begin{itemize}
\item[--] Pions:
\[ U(x)\equiv \xi(x) \xi(x)\ ,\ \ \ \ \xi(x)=e^{i\pi/f_\pi}
       \ ,\ \ \ \  \pi= \mbox{\boldmath$\tau$}
       {\bf\cdot}\mbox{\boldmath$\pi$} /2
\]

\item[--] Nucleons:
\[
N(x)=\left(\!\!\begin{array}{c} p(x) \\[10pt] n(x) \end{array}\!\! \right)
\]

\item[--] Rho:
\[
    \rho_\mu = \mbox{\boldmath$\tau$}
       {\bf\cdot}\mbox{\boldmath$\rho$} /2
\]
\item[--] CCWZ's nonlinear realization of
$SU(2)_{\rm L} \otimes SU(2)_{\rm R}$:

\[
(\xi, \rho_\mu, N)
         \lower 1.4ex \vbox{\hbox{%
   $\ \stackrel{\textstyle \longrightarrow}
    {\scriptstyle L\otimes R } \ $}}
        (\xi', \rho'_\mu, N')
\]

where $L \in SU(2)_{\rm L}$, $R \in SU(2)_{\rm R}$, and
\vspace{.2in}
\begin{eqnarray*}
\xi'(x) &=& L \xi(x) h^{\dagger}(x) = h(x) \xi(x) R^{\dagger}
                         \\[4pt]
\rho'_\mu(x) &=& h(x) \rho_\mu (x) h^{\dagger}(x)
                          \\[4pt]
 N'(x) &=& h(x)N(x)
\end{eqnarray*}

Note
\[
h(x)=h(\mbox{\boldmath$\pi$}(x),L,R) \in SU(2)_{\rm V}
\]
which also guarantees
$\pi$ to be a pseudoscalar under parity:
$\xi
         \lower 1.4ex \vbox{\hbox{%
   $\ \stackrel{\textstyle \longrightarrow}{\scriptstyle P } \ $}}
     \xi^\dagger  $


\end{itemize}
\end{slide}

\begin{slide}       
\vspace*{-1in}
\normalsize
\centerline{$\bullet$ {\bf Gauge Invariance}}
\raggedright
\small
\begin{itemize}
\vspace*{-0.1in}
\item[--] Under $U(1)_{\rm EM}$, the fields transform similarly with

\[
L=R=h(x)=\exp [i\alpha(x)(\tau_3+Y)/2]
\]

$Y=0$ for $\pi$ or $\rho$ and  $Y=1$ for $N\,$.

\item[--]
Allow for the most general gauge invariant terms instead of the
usual minimal substitutions.

\item[--] As a result, universality
of the couplings need
{\bf not} be enforced:
\vspace*{-0.1in}
\[ g\neq g_{\rho\pi\pi} \neq g_\rho \neq g_\gamma \]

where $g_\rho$ is the $\rho N N$ coupling and $g_\gamma$ the rho-photon
coupling (see below).
\end{itemize}

\normalsize
\vspace*{.2in}
\centerline{$\bullet$ {\bf Broken Scale Invariance}}

\small

\begin{itemize}
\item[--]
A light scalar {\it might\/}  be associated with
the broken scale invariance with its potential
constrained by the low-energy
theorems. [~Phys.~Rev.~C52~(1995)1368~]

\item[--] Field redefinitions (see below) make the
constraints weak. Therefore we adopt a general potential
for the scalar $\phi$:
\[
      V_{\rm S}(\phi) = m_{\rm s}^2\phi^2
           \left ({1\over 2}
              +{\kappa_3\over 3!}\,{g_{\rm s}\phi \over M}
             + {\kappa_4 \over 4!}\,{g_{\rm s}^2\phi^2\over M^2}
             +\cdots \right)
\]
\end{itemize}
\end{slide}

\begin{slide}       
\vspace*{-1in}
\normalsize
\centerline{$\bullet$ {\bf NDA and Naturalness}}
\raggedright

\small
\begin{itemize}
\item[--]
NDA rules  for assigning a dimensional coefficient of the right size
to any term [Georgi], with $M=\,$nucleon mass taken to be the generic
cutoff scale:
\begin{enumerate}
\item include a factor of $1/f_\pi$ for each strongly interacting
         field, e.g. $\phi$ or $\omega_\mu$;
\vspace*{-.15in}
\item include an overall factor of $f_\pi^2 M^2$;
\vspace*{-.15in}
\item add factors of $M$ to get the dimension to 4.
\end{enumerate}

\vspace*{-.15in}
\item[--]
Implied ``naturalness'' assumption:
dimensionless coefficients, after extracting the dimensional
factors and some
counting factors, are of $O(1)$.
\vspace*{-.15in}
\item[--]
Assuming NDA, a truncation of the
lagrangian is effective only if
the density is not too high. The reason is that,
for nuclear matter at saturation or at the center of finite nuclei,
the scalar ($\phi$) and vector ($\omega$) mean fields
\[ g_{\rm s}\phi_0/M,\  g_\omega V_0/M \approx  0.3\sim 0.4 \]
and the gradients on the surfaces of finite nuclei
\[g_{\rm s}|\mbox{\boldmath$\nabla$}\phi_0|/M^2,\
g_\omega |\mbox{\boldmath$\nabla$}V_0 |/M^2 \approx (0.2)^2\, .
\]

\vspace*{-.15in}
\item[--] Based on this mean-field estimate, we will count a field
and a derivative to be of the same order.

\vspace*{-.15in}
\item[--]
Fits to the properties of nuclei should determine if
nature
likes the naturalness assumption
and the expansion in powers of the fields and
their derivatives.

\end{itemize}

\end{slide}

\begin{slide}       
\vspace*{-1in}
\normalsize
\centerline{$\bullet$ {\bf Field Redefinitions}}
\raggedright

\vspace*{0.2in}
\small
 \begin{itemize}
\item[--]  Field redefinitions leave the on-shell $S$-matrix
invariant and, among other things, they  are usually used
to remove redundant terms so that there are no terms with
$\partial^2$ acting on
a boson field or $\partial_\mu$ acting on a fermion field
{\it except} in the kinetic terms.

\item[--]  In finite-density applications,
the exchanged non-Goldstone bosons are far
off-shell.
They are introduced to simulate the
intermediate-range interactions; the interaction ranges
remain the same
since the field redefinitions do not
change the quadratic terms.

\item[--] Higher-order derivatives on the non-Goldstone
fields are small
compared to their masses and need not be
eliminated.

\item[--] We use the
field redefinitions to simplify the interaction terms
between the nucleon and the non-Goldstone bosons
such that they are essentially of
the Yukawa form. For example, the following terms are redundant:
\[
    \overline N  N \phi^2\ ,\ \overline N  N\phi^3\ ,\
     \overline N  N\partial^2\phi \ ,\
      \overline N  N   \omega_\mu\omega^\mu  \ ,\, \cdots     \ .
\]
\item[--]
It is still convenient to remove all the
derivatives on the nucleon fields except in the kinetic energy term.

\end{itemize}

\end{slide}

\begin{slide}       
\vspace*{-1.1in}
\normalsize
\centerline{$\bullet$ {\bf Vacuum Effects}}
\raggedright
\small
\begin{itemize}
\item[--]
The QCD vacuum is modified
by interactions with valence nucleons. However,
the vacuum modifications are not adequately treated  in a hadronic
model, so they must be disentangled from valence dynamics and the
effects incorporated into the coefficients of the
effective lagrangian.
\vspace*{-0.1in}
\item[--] Vacuum baryon and non-Goldstone boson
loops are short-distance physics and should be integrated out
so that an expansion in powers of the fields and derivatives is
possible.
But their  fields are still needed in the lagrangian
to account for the valence nucleons and to treat the mean fields
conveniently.
\vspace*{-0.1in}
\item[--] Formally, these loops can be
eliminated by introducing
counterterms, which is not a problem since {\it all possible
interaction terms are already there.}

\end{itemize}

\vspace*{0.2in}
\normalsize
\centerline{{\bf THE LAGRANGIAN}}
\vspace*{0.1in}
\small
Split the lagrangian into the nucleon and meson parts:
\[  {\cal L}= {\cal L}_{\rm N}+{\cal L}_{\rm M}\, . \]
\vspace*{-0.1in}
The nucleon part is
\vspace*{0.2in}
\begin{eqnarray*}
{\cal L}_{\rm N}(x) &=&
         \overline N \Big (i\gamma^{\mu}{\cal D}_\mu
             +g_{\rm \scriptscriptstyle A}\gamma^{\mu}\gamma_5a_{\mu}
              -M +g_{\rm s }\phi \Big )N
                                           \\[3pt]
         & &
       -{f_\rho g_\rho \over 4M}\overline N
        R_{\mu\nu}  \sigma^{\mu\nu} N
         -{f_\omega g_\omega \over 4M}\overline N\omega_{\mu\nu}
                  \sigma^{\mu\nu} N
                              \\[3pt]
       & &
      -{e \over 4M}F_{\mu\nu}
         \overline N \lambda   \sigma^{\mu\nu} N
                                \\[3pt]
        & &
           -{e\over 2M^2}
           \overline{N}\gamma_{\mu}(c_{\rm s}+c_{\rm v}\tau_3)N
               \partial_\nu F^{\mu\nu}
           + \cdots
\end{eqnarray*}

where $\sigma_{\mu\nu}=i[\gamma_{\mu},\gamma_{\nu}]/2\,$,
the nucleon covariant derivative is
\end{slide}

\begin{slide}       
\vspace*{-1.2in}
\raggedright
\small
\[
  {\cal D}_\mu = \partial_{\mu}
            +iv_{\mu}+ig_{\rho}\rho_\mu
             +ig_{\omega }\omega_\mu
             +{i\over 2}eA_\mu(1+\tau_3)
\]
with the vector and axial vector fields defined as
\[v_{\mu} = -{i \over 2}(\xi^{\dagger} \partial_{\mu} \xi +
    \xi \partial_{\mu}\xi^{\dagger} )\, , \ \ \
 a_{\mu} =  -{i \over 2}(\xi^{\dagger} \partial_{\mu} \xi -
    \xi \partial_{\mu}\xi^{\dagger} ) \ ,\]

the field strength tensors are, with
$R_\mu \equiv \rho_\mu + v_\mu / g\,$,
\begin{eqnarray*}
\omega_{\mu\nu}&=&\partial_\mu \omega_\nu -
\partial_\nu \omega_\mu\, ,\ \ \
 F_{\mu\nu}=\partial_\mu A_\nu - \partial_\nu A_\mu\\[2pt]
R_{\mu\nu} & = & \partial_{\mu} R_{\nu} -\partial_{\nu}
      R_{\mu} + i g [\rho_{\mu},{\rho}_{\nu}]
\end{eqnarray*}

and the magnetic-moment operator of the nucleon
\[\lambda ={1\over 2} \lambda_{\rm p} (1+\tau_3)
           + {1\over 2} \lambda_{\rm n} (1-\tau_3) \ ,\]
with $\lambda_{\rm p}=1.79$ and $\lambda_{\rm n}=-1.91$ the
anomalous moments.

\begin{itemize}
\item The $\omega$ meson can be considered as a singlet in $SU(2)$
    symmetry.

\item Terms of next order are
\[
 \overline N\rho_{\mu\nu}
                  \sigma^{\mu\nu} N\phi\, , \
 \overline N\omega_{\mu\nu}
                   \sigma^{\mu\nu} N\phi\, , \  \cdots \]
which are small since
the tensor terms are already small and
the main effects of the above  are slight modifications of
    $f_\rho$ and $f_\omega$ at finite
density due to the nonvanishing scalar expectation value.

\item Nucleon contact terms such as
\[
\overline N N \overline N N\, , \ \
\overline N \gamma_\mu N \overline N \gamma^\mu N\, ,\  \cdots
\]
have been taken into account by fitting $g_{\rm s}$ and
$g_\omega$. In other words, the dominate effects of other
particles in the same scalar and vector channels are implicit in
$g_{\rm s}$ and
$g_\omega$.

\end{itemize}

\end{slide}

\begin{slide}       
\vspace*{-1.1in}
\raggedright
\small
Up to quartic order, the meson part of $\cal L$ is
\begin{eqnarray*}
{\cal L}_{\rm M}(x) &=&
       -{1\over 2}{\rm tr}\, (R_{\mu\nu}R^{\mu\nu})
                   -{1 \over 4}\omega_{\mu\nu}\omega^{\mu\nu}
          + {1\over 2}\partial_\mu \phi \partial^\mu \phi
                                           \\[3pt]
         &  &
           -{1 \over 4}F_{\mu\nu}F^{\mu\nu}
         +{f_{\pi}^2\over 4}\, {\rm tr}\,
                  (\partial _{\mu}U\partial ^{\mu}U^{\dagger})
                        \\[3pt]
       &  &  \null
      -2 f_{\pi}^2 eA^{\mu}{\rm tr}\,(v_{\mu}\tau_3)
          -{e F_{\mu\nu} \over 2g_{\gamma}}\,\Big[\,
           {\rm tr}\,(\tau_3 R^{\mu\nu})
            +{1\over 3}\omega^{\mu\nu} \Big]    \\[3pt]
        &  &  \null
       + {1\over 2}\bigg (
      1+ \eta_1 {g_{\rm s}\phi\over M}
        + {\eta_2\over 2} {g_{\rm s}^2\phi^2\over M^2} \bigg )
                   m_{\omega}^2 \omega_{\mu}\omega^{\mu}
                        \\[3pt]
        &  &  \null
         +{1\over 4!}\zeta
             g_\omega^2 (\omega_{\mu}\omega^{\mu})^2
         +   \bigg ( 1+ \eta_\rho {g_{\rm s}\phi\over M} \bigg )
             m_\rho^2 \,{\rm tr}\,(\rho_{\mu}\rho^\mu)
                         \\[3pt]
         &  &       \null
            - m_{\rm s}^2\phi^2\bigg ({1\over 2}+{\kappa_3\over 3!}\,
            {g_{\rm s}\phi \over M} + {\kappa_4 \over 4!}\,
           {g_{\rm s}^2\phi^2\over  M^2}\bigg ) + \cdots
\end{eqnarray*}
\vspace*{-.15in}
\begin{itemize}
\item Apart from some conventional definitions,
the couplings have been defined so that they should be of
order unity  if the NDA is valid.
\vspace*{-.15in}
\item Some unimportant terms, such as those with two photon and
two rhos or higher powers of rhos, have been omitted.
\vspace*{-.15in}
\item The numerical importance of higher-order
terms is checked by including in $\cal L$  these fifth-order terms:
\begin{eqnarray*}
    {\cal L}_5 &=&  {1 \over 4}\, {g_{\rm s}\phi\over M} [
          2\alpha_1  (\partial_\mu \phi)^2
            -\alpha_2  \omega_{\mu\nu}^2 ]
                  - {\kappa_5\over 5!}\,
                {g_{\rm s}^3\phi^3\over M^3} m_{\rm s}^2\phi^2
                                \\[3pt]
         &  &       \null
          + {\eta_3\over 3!}\,
                {g_{\rm s}^3\phi^3\over M^3}
               \cdot {1\over 2} m_{\omega}^2 \omega_{\mu}\omega^{\mu}
                         + {\zeta_1 \over 4!}\,
                   {g_{\rm s}\phi\over M}
             g_{\omega}^2 (\omega_{\mu}\omega^{\mu})^2
\end{eqnarray*}

\end{itemize}

\end{slide}

\begin{slide}       
\vspace*{-.8in}
\raggedright
\centerline{{\bf EM STRUCTURE OF THE NUCLEON}}

\small
The electromagnetic current is a sum of point-nucleon and meson-cloud
contributions (pion terms omitted):
\begin{eqnarray*}
 J_\mu &=&J_\mu^{\rm PT}+J_\mu^{\rm M}\\[2pt]
J_\mu^{\rm PT}&=&{1\over 2}\overline{N}(1+\tau_3)\gamma_\mu N
       +{1\over 2M}\partial_\nu(\overline{N}\lambda\sigma^{\mu\nu}N)
         \\[2pt]
    & &\ \ \ \ \  -{1\over 2M^2}\partial^2[\overline{N}
           (c_{\rm s}+c_{\rm v}\tau_3)N]  \\[2pt]
J_\mu^{\rm M} &=&
         {1\over g_\gamma}(\partial_\nu \rho_3^{\mu\nu}
             +{1\over 3} \partial_\nu\omega^{\mu\nu})
         +2f_\pi^2 \, {\rm tr}\, (v_\mu \tau_3)
\end{eqnarray*}
from which the EM isoscalar and isovector Dirac and Pauli
form factors of the nucleon can be read off:
\begin{eqnarray*}
F_1^{\rm s}(Q^2) &=& {1\over 2} - {c_{\rm s}\over 2}\,{Q^2 \over M^2}
     -{g_\omega\over 3g_\gamma}\,
       {Q^2 \over Q^2+m_\omega^2}+\cdots     \\[3pt]
F_1^{\rm v}(Q^2) &=& {1\over 2} - {c_{\rm v}\over 2}\,{Q^2 \over M^2}
     -{g_\rho\over 2g_\gamma}\,
       {Q^2 \over Q^2+m_\rho^2}+\cdots   \\[3pt]
F_2^{\rm s}(Q^2) &=& {\lambda_p+\lambda_n\over 2}
     -{f_\omega g_\omega\over 3g_\gamma}\,
       {Q^2 \over Q^2+m_\omega^2}+\cdots    \\[3pt]
F_2^{\rm v}(Q^2) &=& {\lambda_p-\lambda_n\over 2}
     -{f_\rho g_\rho\over 2g_\gamma}\,
       {Q^2 \over Q^2+m_\rho^2}+\cdots
\end{eqnarray*}
\vspace{-.1in}
\begin{itemize}
\item The photon {\bf does} see an extended nucleon; there is no need
for external ``intrinsic'' form factors.
\vspace*{-.15in}
\item The charge density of a nucleus,
$ \rho_{\rm ch} = \rho^{\rm PT}+\rho^{\rm M} $,
has contributions from the neutrons as well as from the meson clouds
which, as seen below, smear out the point nucleon charge oscillations
and increase the RMS radius.
\end{itemize}
\end{slide}

\begin{slide}       
\vspace*{-1.1in}
\raggedright
\centerline{{\bf ONE-LOOP RESULTS}}
\small


\begin{itemize}

\item A systematic calculation to quartic order requires
      at least the inclusion of
      two-loop contributions.
\item Calculate only through  one loop in this first investigation.

\item Parameters are optimized to fit the following observables
in $^{16}$O,  $^{88}$Sr, and $^{208}$Pb
(using a $\chi^2$ minimization):

\begin{quote}
binding energies per nucleon $E_{\rm B}/A$,
\\ rms charge radii $\langle r^2 \rangle^{1\over 2}$,\\
spin-orbit splittings $\Delta E_{\rm SO}$ of
least-bound\\ \hspace*{.2in} protons and neutrons,\\
diffraction minimum
sharp  radii $R_{\rm dms}$; \\
 also the $1h_{9/2}$
energy level
 of  $^{208}$Pb.
\end{quote}

Here $R_{\rm dms}\equiv 4.493/Q_0^{(1)}$, with
$Q_0^{(1)}$ the first zero of the charge form factor of the
nucleus in momentum space. The $1h_{9/2}$ level is used
to fix the overall scale of the $^{208}$Pb energy levels,
which is strongly affected by $g_\rho$.

\end{itemize}
\end{slide}

\begin{slide}

$\bullet$  Parameters:

\small
\renewcommand{\arraystretch}{1.6}
\setlength{\tabcolsep}{.07in}
\begin{center}
\begin{small}
\begin{tabular}{|ccccccccc|}  \hline\hline\hline
\multicolumn{9}{|c|}{Input Parameters
     (mass in MeV and radius in fm)}\\
\hline
$M$ & $m_{\omega}$ & $m_{\rho}$ & $\lambda_p$ &
$\lambda_n$& $g_\gamma$ & $ \langle r^2 \rangle_{{\rm s}1}^{1\over2}$
& $\langle r^2 \rangle_{{\rm v}1}^{1\over2}$
& $\langle r^2 \rangle_{{\rm v}2}^{1\over2}$
\\       \hline
939 &  782 & 770 & 1.79 & -1.91 &5.01& 0.79 & 0.79 & 0.88  \\
\hline
\end{tabular}
\end{small}
\end{center}

\vspace{0.1in}
$g_\gamma$ is determined from
$\Gamma_{\rho^0\rightarrow e^+e^-}=6.8\,$keV
\vspace{0.1in}

\renewcommand{\arraystretch}{1.4}
\setlength{\tabcolsep}{.06in}
\begin{center}
\begin{small}
\begin{tabular}{|cccccccc|}  \hline\hline\hline
\multicolumn{8}{|c|}{Fitted Parameters }\\
\hline
$g_{\rm s}/4\pi$ & $g_{\omega}/4\pi$ & $g_{\rho}/4\pi$
& $\eta_1$& $\eta_2$& $\zeta$ & $\kappa_3$ & $\kappa_4$
\\       \hline\hline
\tiny
$0.78$ &\tiny $0.90$ &\tiny $0.55$ &\tiny $-0.71$ &\tiny $1.39$
  &\tiny $1.20$  &\tiny $-0.33$ &\tiny $1.79$ \\
\hline
%
\hline
 $m_{\rm s}/M$ & $\eta_\rho$ &  $f_{\omega}/4$ &
 $\eta_3$ & $\kappa_5$ & $\zeta_1$ & $\alpha_1$ & $\alpha_2$
 \\       \hline
 \tiny $0.58$ & \tiny $-1.11$ & \tiny $0.024$ &\tiny $-0.24$
&\tiny $2.98$ &\tiny $0.52$ &\tiny $0.49$ &\tiny $0.002$
\\ \hline
\end{tabular}
\end{small}
\end{center}

\vspace{0.1in}

According to NDA, all entries  are natural if $O(1)$. Also
we have
\begin{eqnarray*}
 c_{\rm s} &=& {M^2 \over 6}  \langle r^2 \rangle_{{\rm s}1}
              - {2\over 3}\,{g_\omega\over g_\gamma}\,
                {M^2\over m_\omega^2} =  0.19 \\[6pt]
 c_{\rm v} &=& {M^2 \over 6}  \langle r^2 \rangle_{{\rm v}1}
              - {g_\rho\over g_\gamma}\,
                {M^2\over m_\rho^2} = 0.31 \\[6pt]
 {f_\rho\over 4} &=& {\lambda_p  - \lambda_n \over 24}\,
           {g_\gamma \over g_\rho}\,m_\rho^2\, \langle r^2
            \rangle_{{\rm v}2} =  1.33
\end{eqnarray*}

\end{slide}

\begin{slide}       
\vspace*{-1.in}
\raggedright

$\bullet$ Results {\small (including center-of-mass corrections)}

\small

\renewcommand{\arraystretch}{1.5}
\setlength{\tabcolsep}{.08in}
\begin{center}
\begin{small}
\begin{tabular}{|l|ccc|ccc|cc|}  \hline\hline\hline
\multicolumn{1}{|c}{ }  &
\multicolumn{3}{|c|}{$E_{\rm B}/A$ (MeV)} &
\multicolumn{3}{|c|}{$\langle r^2 \rangle^{1\over 2}_{\rm chg}$ (fm)}&
\multicolumn{2}{|c|}{$R_{\rm dms}$ (fm)} \\
\hline
   & PC$^*$ & here & expt& PC & here & expt&  here & expt
\\ \hline
$^{16}{\rm O}$   &  7.97 &  7.98    & 7.98
                 &  2.73 &  2.70    & 2.74
                 &  2.78 &  2.76
\\       \hline
$^{88}{\rm Sr}$  &  8.75 &  8.71    & 8.73
                 &  4.21 &  4.18    & 4.20
                 &  4.98     & 4.99
\\       \hline
$^{208}{\rm Pb}$ &  7.87 &  7.89    & 7.87
                 &  5.51 &  5.49    & 5.50
                 &  6.79 &  6.78
\\       \hline
\end{tabular}
\end{small}
\end{center}
* Point-coupling model of Nikolaus et al.


\renewcommand{\arraystretch}{1.5}
\setlength{\tabcolsep}{.08in}
\begin{center}
\begin{small}
\begin{tabular}{|l|ccc|ccc|}  \hline\hline\hline
  &
\multicolumn{3}{|c|}{ $\Delta E_{\rm SO}^{(n)}$ (MeV)}
 &
\multicolumn{3}{|c|}{ $\Delta E_{\rm SO}^{(p)}$ (MeV)}
\\ \hline
  & PC & here & expt& PC & here & expt
\\ \hline
$^{16}{\rm O}$   & 6.4 & 5.8  &  6.2
                 & 6.4 & 5.8  &  6.3
\\       \hline
$^{88}{\rm Sr}$  & 1.9 & 2.0  &  (1.5)$^{**}$
                 & 6.1 & 5.7  &  (3.5)$^{**}$
\\       \hline
$^{208}{\rm Pb}$ & 0.91 & 0.77 &  0.90
                 & 1.96 & 1.58  &  1.33
\\       \hline
\end{tabular}
\end{small}
\end{center}
$**$ calculated from Bohr and Mottelson:
\[
 \Delta E_{\rm SO}^{(n)} =\Delta E_{\rm SO}^{(p)} \approx
   10(2l+1)/A^{2/3} \ {\rm MeV}
\]

-- The {\bf predicted} saturation properties of nuclear matter
with
$k_{\rm F}$ the Fermi momentum in fm$^{-1}$,
$K$ the compressibility,
$a_4$ the symmetry energy, and $M^*$ effective mass. Also given is
the fifth-order contribution ($\Delta E_5$) to $E_{\rm B}/A$.
The mass unit is MeV.

\renewcommand{\arraystretch}{1.5}
\setlength{\tabcolsep}{.1in}
\begin{center}
\begin{small}
\begin{tabular}{|cccccc|c|}  \hline
$E_{\rm B}/A$ & $k_{\rm F}$ & $K$ & $a_4$ & $M^*/M$ & $g_{\omega}V_0$
& $\Delta E_5$\\
\hline
\tiny (MeV) & (fm$^{-1}$) & \tiny (MeV) & \tiny (MeV) &  &  \tiny
(MeV)  &  \tiny (MeV)  \\
\hline
 16.1 & 1.31 & 214. & 37.5 & 0.622 & 284 & 0.45 \\
\hline
\end{tabular}
\end{small}
\end{center}

\end{slide}

\begin{slide}       
\vspace*{-1.in}
\raggedright
\small

\begin{minipage}{7.5in}
 \setlength{\epsfxsize}{3.3in}
 \begin{minipage}{3.35in}
  \epsfbox{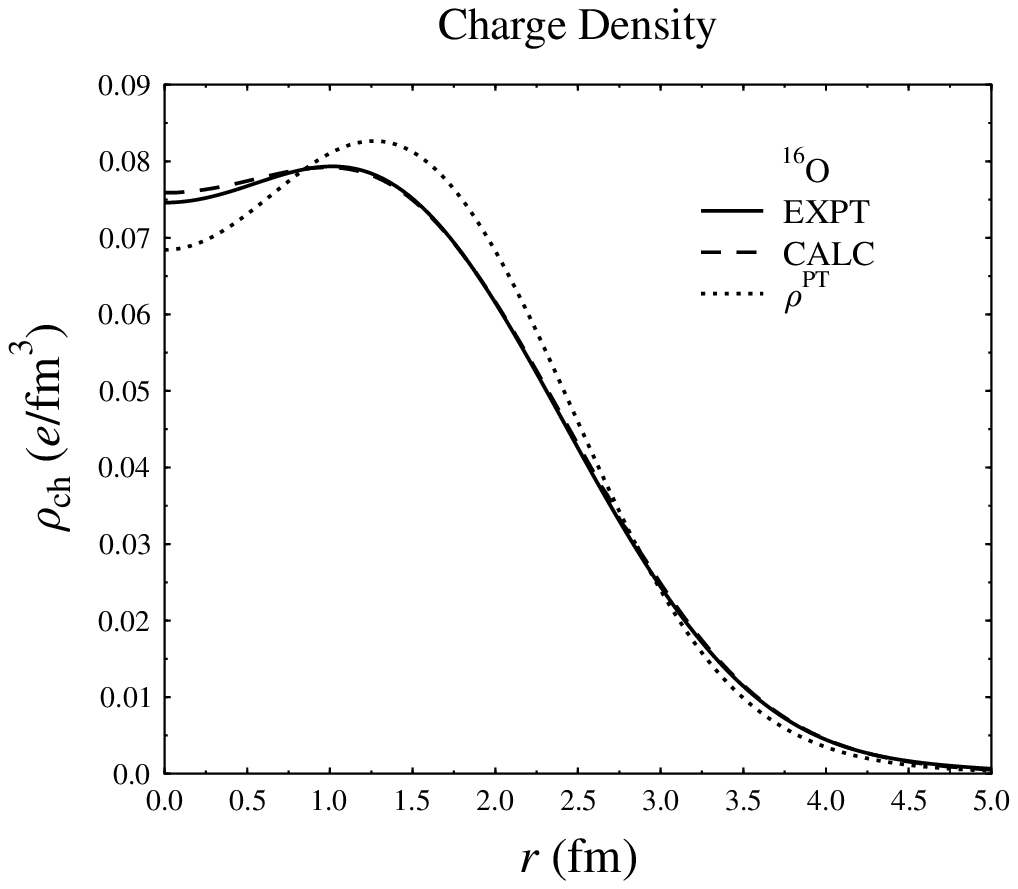}
 \end{minipage}
 \setlength{\epsfxsize}{3.2in}
 \begin{minipage}{3.25in}
  \epsfbox{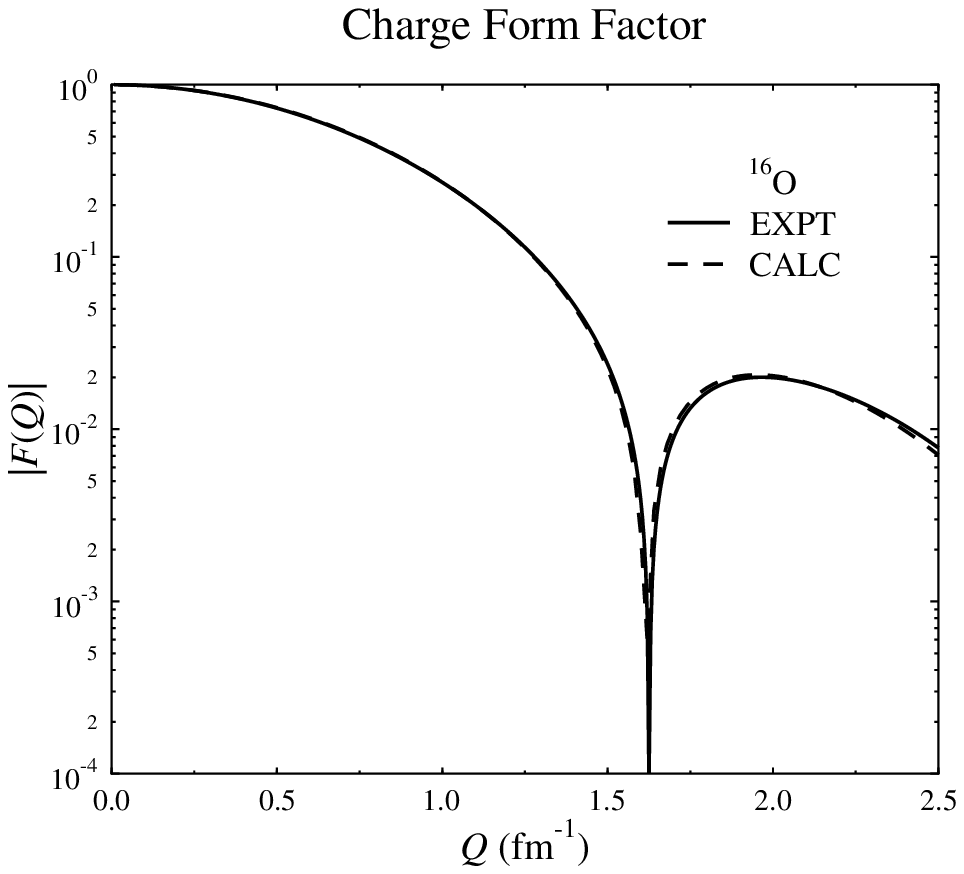}
 \end{minipage}
\end{minipage}

\vspace*{20pt}

\begin{minipage}{7.5in}
 \setlength{\epsfxsize}{3.3in}
 \begin{minipage}{3.35in}
  \epsfbox{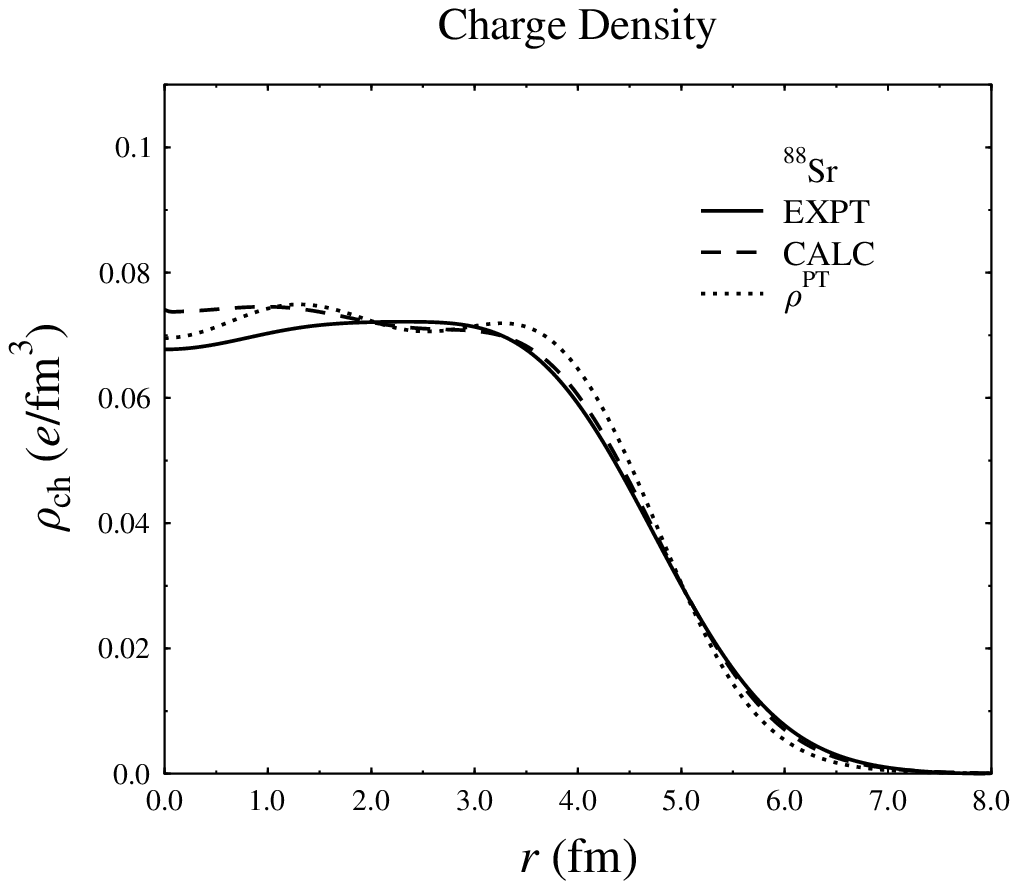}
%
 \end{minipage}
 \setlength{\epsfxsize}{3.2in}
 \begin{minipage}{3.25in}
  \epsfbox{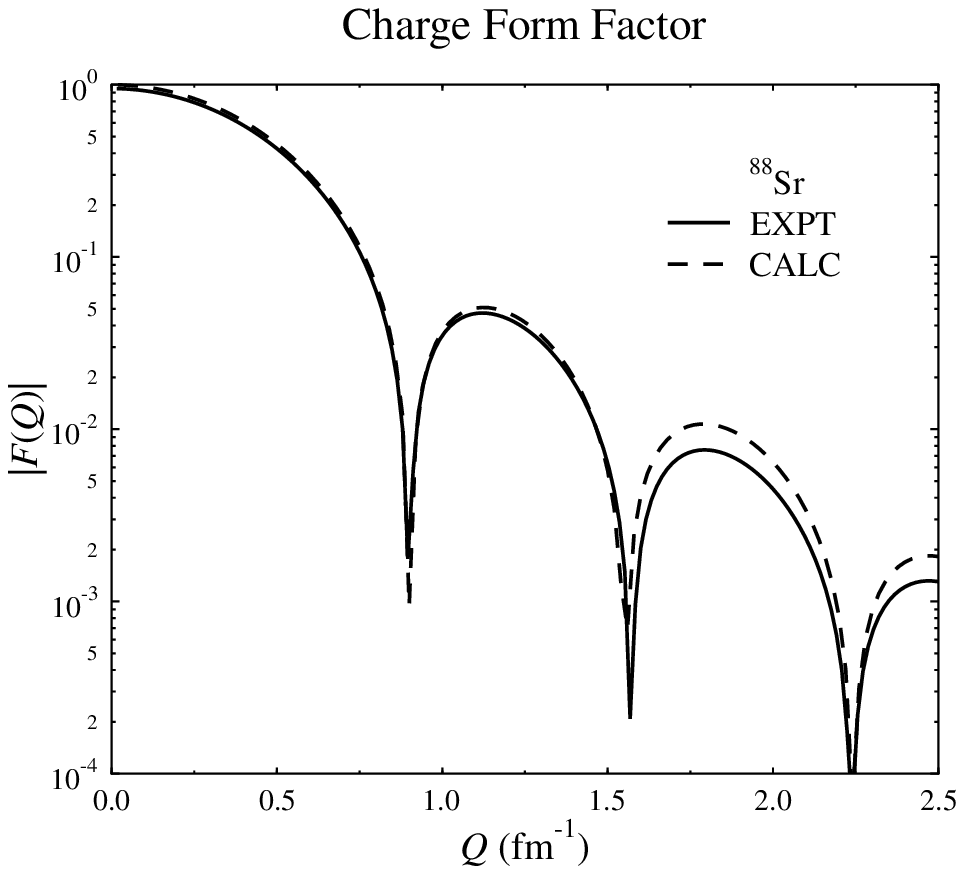}
 \end{minipage}
\end{minipage}

\vspace{.2in}
Note the agreement between the calculated and experimental results
at low momenta.
\end{slide}

\begin{slide}       
\vspace*{-1.in}
\raggedright
\small

\begin{minipage}{7.5in}
 \setlength{\epsfxsize}{3.3in}
 \begin{minipage}{3.3in}
  \epsfbox{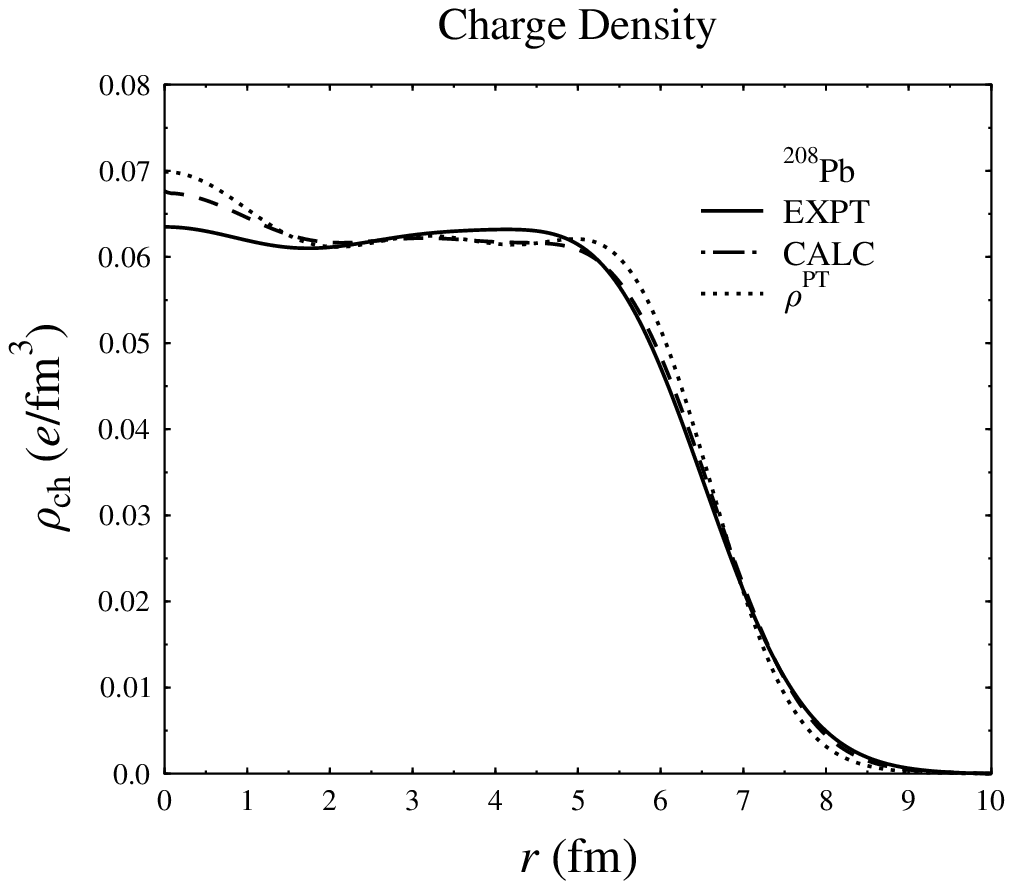}
 \end{minipage}
 \setlength{\epsfxsize}{3.2in}
 \begin{minipage}{3.2in}
  \epsfbox{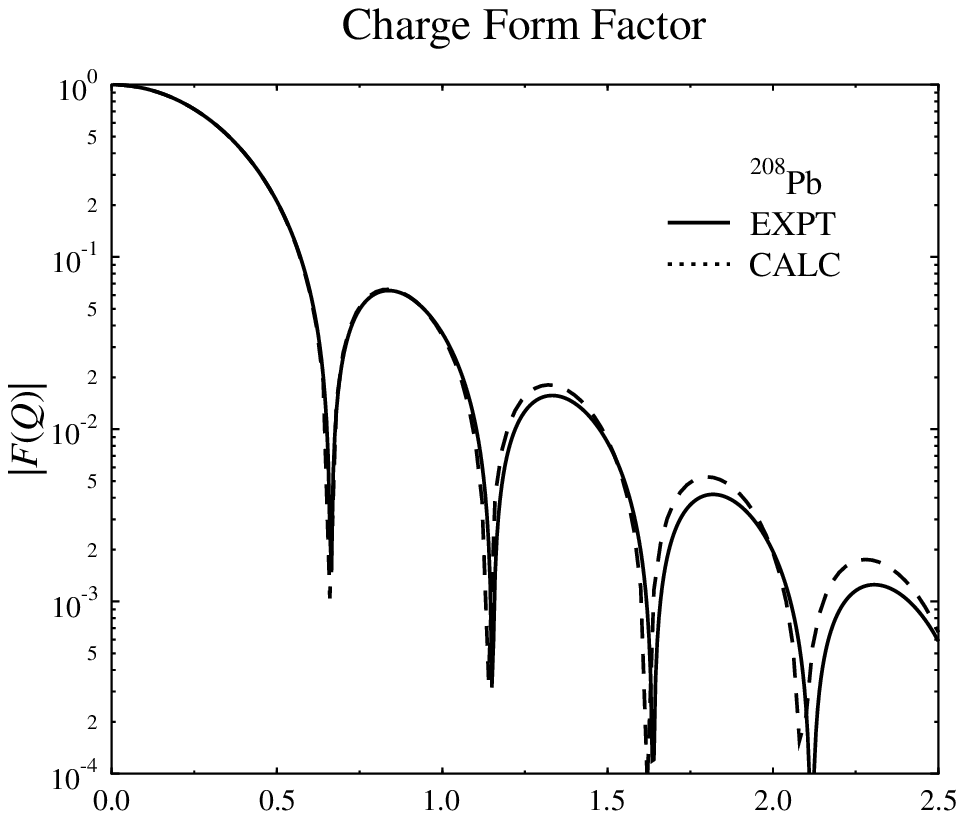}
 \end{minipage}
\end{minipage}

\vspace*{0.8in}

\begin{center}
\begin{minipage}{2.2in}
 \setlength{\epsfxsize}{2.2in}
  \epsfbox{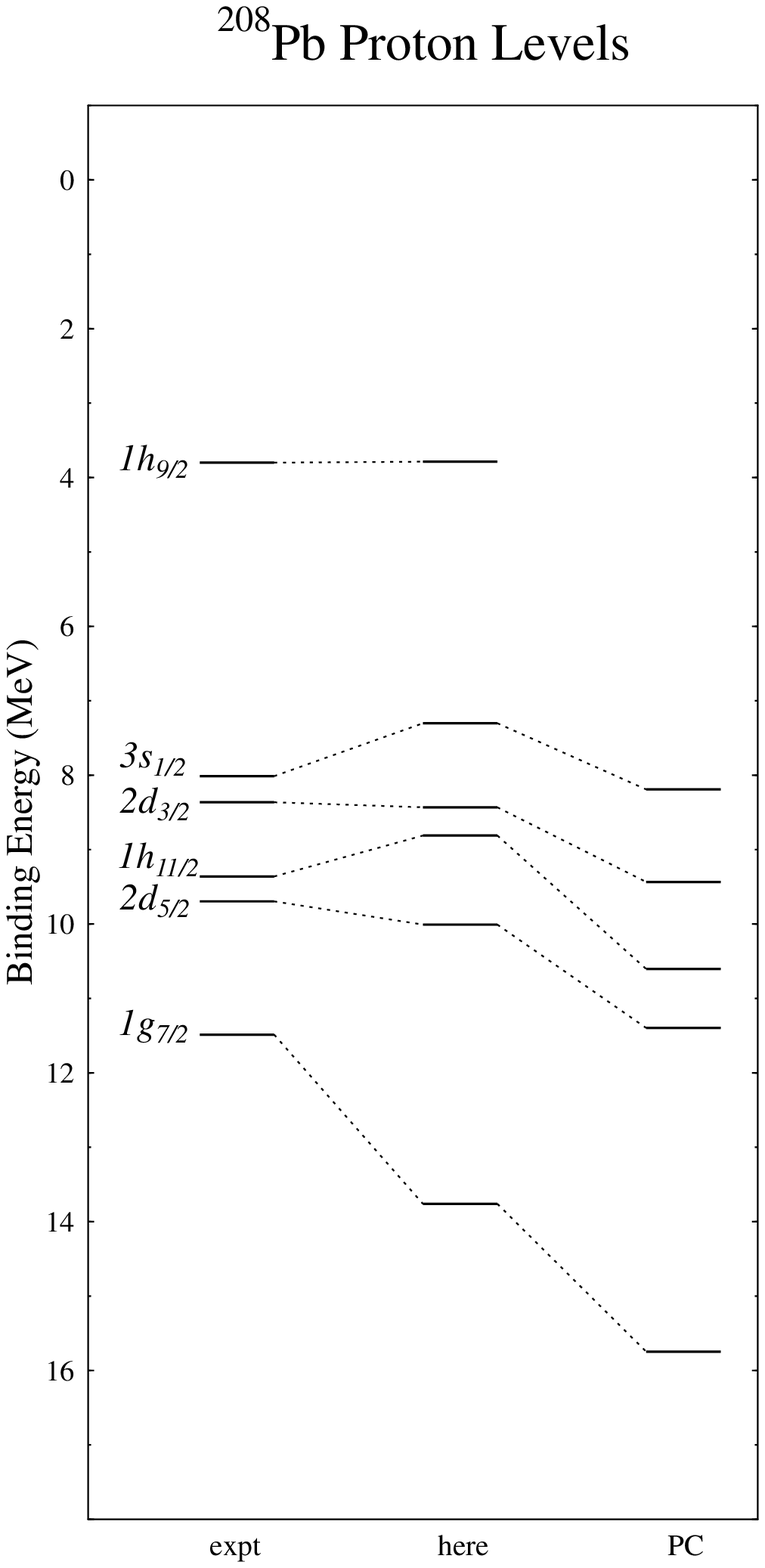}
\end{minipage}
\end{center}
\end{slide}

\begin{slide}       
\vspace*{-1.in}
\centerline{{\bf SUMMARY AND FUTURE WORK}}
\raggedright
\small

\begin{itemize}
\item The low-energy structure of the nucleon and
      finite nuclei can be well-described
      by a chiral effective theory of hadrons at one-loop order.
\item The NDA and naturalness assumption are indeed compatible with the
      observed properties of nuclei. The fit parameters are natural and
      the results are not driven by the last terms kept.
\item Descriptions of the spin-orbit splittings of the least bound
      nucleons and the shell structure are generally good, although
      there are some discrepancies in the fine details of the energy
      levels of heavy nuclei.
\item The meson clouds play an important role in smearing out the
      point nucleon charge fluctuations.
\item Future work:
   \begin{itemize}
      \item Two-loop calculations (in progress);
      \item Connections with point-coupling lagrangian;
      \item Role of the explicit $\Delta$ degrees of freedom.
   \end{itemize}

\end{itemize}

[{\em Note added: A journal article  with greater detail and
references will be available soon. }]

\end{slide}

\end{document}